# Nanostructured Melt-Spun Sm(Co,Fe,Zr,B)$_{7.5}$ Alloys for High-Temperature Magnets

Sofoklis S. Makridis, George Litsardakis, Kostas G. Efthimiadis, George Papathanasiou, Ioannis Panagiotopoulos, Sabine Höfinger, Josef Fidler, George C. Hadjipanayis, and Dimitris Niarchos

*Abstract*—High coercivity, the highest for Cu-free 2 : 17 Sm–Co ribbons, has been obtained in as-spun ($H_c$ = 21.1 kOe) and short time annealed ($H_c$ = 23.2 kOe) samples of Sm(Co$_{bal}$Fe$_z$Zr$_y$B$_x$)$_{7.5}$ alloys, with varying B, Zr, and Fe content ($x$ = 0–0.06, $y$ = 0–0.16, $z$ = 0.08–0.3) and wheel speed. In as-spun samples, the TbCu$_7$ type structure and in annealed samples the Th$_2$Zn$_{17}$ and CaCu$_5$ type structures is observed, plus fcc Co as minority phase is observed. Reduced remanence ($M_r/M_s$) is higher than 0.7. High-temperature magnetic measurements show very good stability above 300 °C with coercive field as high as 5.2 kOe at 330 °C. For annealed Sm(Co$_{bal}$Fe$_{0.3}$Zr$_{0.02}$B$_{0.04}$)$_{7.5}$, very good loop squareness and high maximum energy product of 10.7 MGOe have been obtained. Increasing Zr content results in less uniform microstructure of annealed ribbons.

## I. INTRODUCTION

PERMANENT magnets based on Sm(Co,Fe,Cu,Zr)$_z$ bulk compounds that are suitable for high-temperature applications above 400 °C are obtained after long and complicated heat treatment, which is required for the development of the characteristic, high coercivity, cellular-lamellar microstructure [1], [2]. In alloys with similar composition, nonequilibrium processing techniques, like rapid quenching or mechanical alloying, have attracted much attention as alternative routes to fabrication of nanostructured, high-temperature magnets [3], [4]. We have recently shown that small boron addition can facilitate the development of high coercivity in nanostructured melt-spun ribbons [5], while the presence of Cu is not a requisite for the development of coercivity and temperature stability [6].

In this paper, we report details on the synthesis, structural characterization, and magnetic properties of Cu-free Sm(Co$_{bal}$ Fe$_z$Zr$_y$B$_x$)$_{7.5}$ melt spun ribbons ($x$ = 0–0.06, $y$ = 0–0.16, $z$ = 0.08–0.3), in which high coercivity is developed in as-spun and short time annealed samples.

## II. EXPERIMENTAL PROCEDURE

Samples were prepared by arc melting under argon atmosphere with an excess of 5%–10% Sm to compensate weight losses during processing. Ribbons have been rapidly solidified from master alloys by melt-spinning, on a copper roller with wheel speed 10 to 70 m/s. Before the annealing process, ribbons were wrapped in tantalum foil and sealed in a quartz tubes under high vacuum ($10^{-5}$ torr) or pure argon ($\sim$1 atm), to avoid oxidization. Crystal structures of as-spun and annealed ribbons were investigated by X-ray diffraction with Fe–K$_\alpha$ radiation and by electron microscopy. Microstructure was examined using a JEOL (JEM-2000FX) transmission electron microscope (TEM). Magnetic properties of the samples were determined by means of an extraction magnetometer with $H_{max}$ = 70 kOe, a superconducting quantum interference device (SQUID) magnetometer with $H_{max}$ = 52 kOe, and a vibrating sample magnetometer (VSM) for tracing thermomagnetic curves of nonmagnetized as-spun ribbons under external field of 300 Oe.

## III. RESULTS AND DISCUSSION

The as-spun ribbons have the TbCu$_7$-type of structure (S.G. P6/mmm), as shown in Fig. 1. The lattice parameters are $a$ = $b$ = 0.495–0.499 nm and $c$ = 0.406–0.411 nm, depending on the composition. The broad peaks in X-ray diffraction patterns of ribbons spun above 30 m/s of the wheel speed indicate a nanoscale crystallite size.

Annealing at 600 °C for 30–75 min results in grain growth without structural changes. Phase transformations are observed after annealing above 800 °C for 1 h (Fig. 2). The Th$_2$Zn$_{17}$ and the CaCu$_5$ types of structure are formed and fcc–Co as minority phase can be observed. The thermomagnetic curves of as-spun ribbons exhibit the transformation of the 1 : 7 phase, as an increase of magnetization around 500 °C–700 °C, the Curie temperature of the remaining 1 : 7 phase, close to 750 °C.

Manuscript received January 2, 2003. This work was supported in part by the Defense Advanced Research Projects Agency (DARPA) Metamaterial Program (USA) and in part by the European Commission (EC) project HITEMAG (G5RD-CT2000-00213).

S. S. Makridis is with the Department of Electrical and Computer Engineering, Aristotle University, Thessaloniki 54124, Greece and also with the Institute of Materials Science, NCSR "Demokritos," Athens 15310, Greece (e-mail: sofmak@eng.auth.gr).

G. Litsardakis is with the Department of Electrical and Computer Engineering, Aristotle University, Thessaloniki 54124, Greece (e-mail: Lits@eng.auth.gr).

K. G. Efthimiadis is with the Department of Physics, Aristotle University, Thessaloniki 54124, Greece (e-mail: kge@auth.gr).

G. Papathanasiou is with the Department of Electrical and Computer Engineering, Democritus University of Thrace, Xanthi 67100, Greece (e-mail: gpap@ee.duth.gr).

I. Panagiotopoulos is with the Department of Materials Science and Engineering, University of Ioannina, Ioannina 45110, Greece (e-mail: ipanagio@cc.uoi.gr).

S. Höfinger and J. Fidler are with the Institute of Applied and Technical Physics, Vienna University of Technology, Vienna A-1040, Austria (e-mail: sabine.hoefinger@ifp.tuwien.ac.at; fidler@tuwien.ac.at).

G. C. Hadjipanayis is with the Department of Physics and Astronomy, University of Delaware, Newark, DE 19716 USA (e-mail: hadji@udel.edu).

D. Niarchos is with the Institute of Materials Science, NCSR "Demokritos," Athens 15310, Greece (e-mail: dniarchos@ims.demokritos.gr).

Digital Object Identifier 10.1109/TMAG.2003.815731

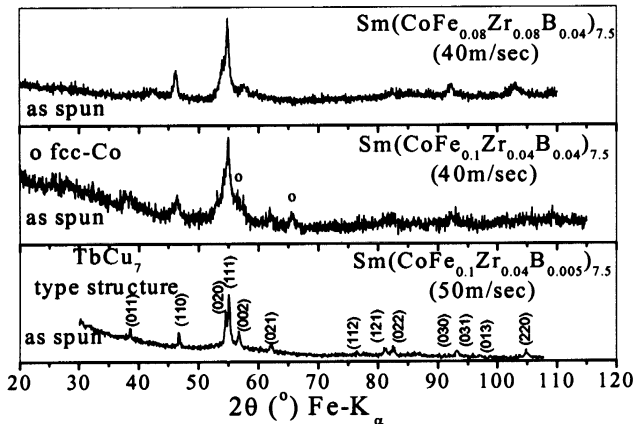

Fig. 1. X-ray patterns of as-spun ribbons.

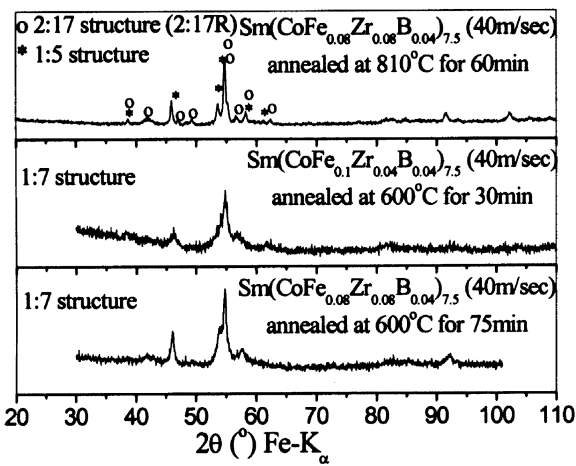

Fig. 2. X-ray diffraction patterns of annealed ribbons.

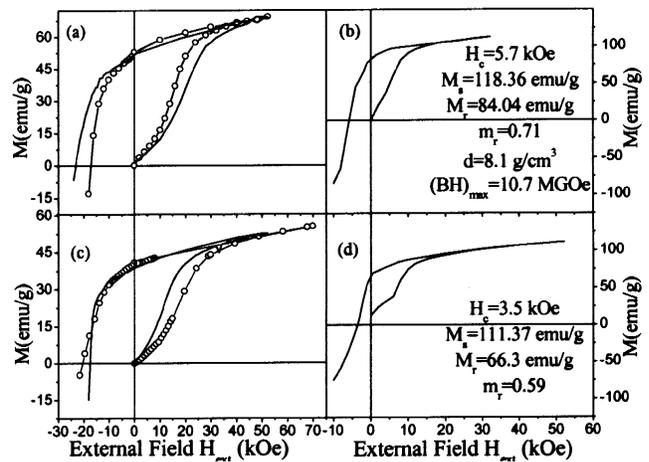

Fig. 3. Hysteresis graphs and properties of annealed ribbons (a) -o-Sm(CoFe$_{0.1}$Zr$_{0.04}$B$_{0.04}$)$_{7.5}$ (40 m/s) annealed at 600 °C for 75 min, — Sm(CoFe$_{0.1}$Zr$_{0.04}$B$_{0.04}$)$_{7.5}$ (60 m/s) annealed at 600 °C for 30 min. (b) Sm(CoFe$_{0.3}$Zr$_{0.02}$B$_{0.04}$)$_{7.5}$ (40 m/s) annealed at 600 °C for 75 min. (c) -o- Sm(CoFe$_{0.08}$Zr$_{0.04}$B$_{0.04}$)$_{7.5}$ (40 m/s) annealed at 600 °C for 75 min and — annealed at 810 °C for 60 min. (d) Sm(CoFe$_{0.1}$B$_{0.04}$)$_{7.5}$ (38 m/s) annealed at 600 °C for 75 min.

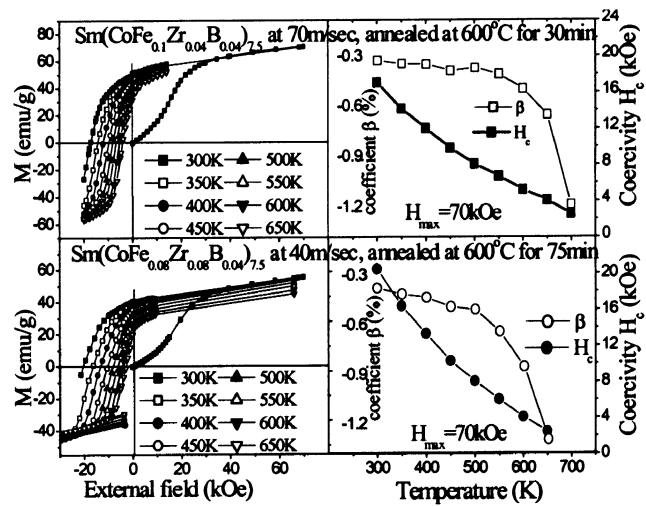

Fig. 4. Loop shape, coercivity dependence with temperature and temperature coefficient of coercivity $\beta$.

The coercivity of as-spun ribbons varies with the roller's orbital wheel speed; it is higher in the range 25–60 m/s with a maximum of ~8 kOe at 50 m/s for $x = 0.015$. In boron-free ribbons, annealed at 700 °C for 40 min, intrinsic coercivity $_iH_c$ is 5 kOe. In boron-substituted samples, annealed at 600 °C for 30–75 min, coercivity varies from 6 to 23.2 kOe, depending on the composition, while for Zr-free Sm(Co$_{bal}$Fe$_{0.1}$B$_{0.04}$)$_{7.5}$, it is only 3.5 kOe.

A small step that appears at low fields in the demagnetization curves of as-spun ribbons is attributed to the existence of noncoupled grains of soft magnetic phase (fcc–Co). After annealing, the loop squareness is improved and the soft step in most cases disappears.

Very good loop squareness and a maximum energy product of $(BH)_{max} = 10.7$ MGOe has been obtained at room temperature for sample Sm(Co$_{bal}$Fe$_{0.3}$Zr$_{0.02}$B$_{0.04}$)$_{7.5}$ after annealing at 600 °C for 75 min, as shown in Fig. 3. In spite of the low value of coercive field (16.1 kOe), the energy product is high because of the high magnetization, as a result of the high Fe and low Zr content.

Remanence enhancement is observed ($M_r/M_s$ is higher than 0.7), as a result of strong exchange coupling between hard–hard or hard–soft nanograins. The nature of magnetic interactions between the grains has been examined by means of $\delta M$ plots of remanence curves for as-spun and annealed ribbons. Higher loop squareness is associated to higher maximum value of $\delta M$ at the coercive field value, which originates from the strong exchange coupling between nanograins. $\delta M_{max}$ is 0.25 for the as-spun sample Sm(CoFe$_{0.04}$Zr$_{0.04}$B$_{0.04}$)$_{7.5}$ ($H_c \sim 8$ kOe) while it is 0.73 for sample Sm(CoFe$_{0.08}$Zr$_{0.08}$B$_{0.04}$)$_{7.5}$ annealed at 810 °C for 60 min ($H_c \sim 17$ kOe). A negative value of $\delta M_{max}$ ($-0.08$) for the annealed sample is associated with weaker dipolar interactions after the magnetization reversal. Recoil loops of the demagnetization curve display high reversibility, especially at fields before the coercive field.

For short time annealed ribbons, coercivity at high temperatures is retained as high as 5.2 kOe above 600 K (330 °C)—a value observed for the first time in Cu-free nanostructured ribbons, as shown in Fig. 4. Samples with Zr atomic content 4%–8% display better thermal stability. In Fig. 4, it is shown that the temperature coefficient of coercivity $\beta$ $(=\Delta_i H_c/(\Delta T \cdot _iH_c) \times 100)$ is much more stable in the range of 300 to 550 K for sample Sm(Co$_{bal}$Fe$_{0.1}$Zr$_{0.04}$B$_{0.04}$)$_{7.5}$ which is annealed at 600 °C for 30 min. This probably results from the more uniform microstructure compared to that of the annealed sample with 0.08% Zr.

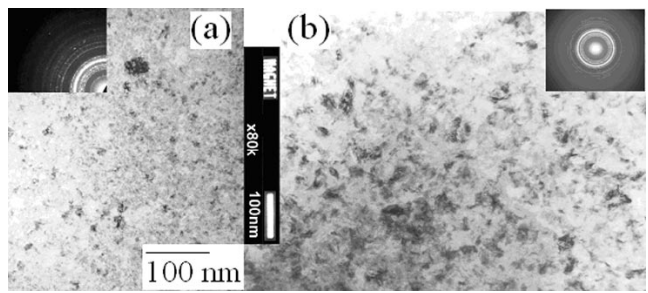

Fig. 5. TEM pictures of sample Sm(CoFe$_{0.1}$Zr$_{0.04}$B$_{0.04}$)$_{7.5}$ (40 m/s) (a) as-spun and (b) annealed ribbons at 600 °C for 30 min.

TEM studies have been used to investigate microstructure and grain size distribution. In diffraction patterns 1 : 7 and fcc-Co have been identified. Average grain size is 30–50 nm in as-spun ribbons and 70–100 nm for annealed ribbons. The grain distribution is very uniform in the scale of nanometers for the samples with square loop (Fig. 5).

Comparing sample Sm(CoFe$_{0.1}$Zr$_{0.04}$B$_{0.04}$)$_{7.5}$ to Sm(CoFe$_{0.08}$Zr$_{0.08}$B$_{0.04}$)$_{7.5}$, we observed that both samples have similar microstructures; it is the one with the finer microstructure (Zr = 0.04) that has higher magnetization and lower thermal variation of coercivity.

Electron microscopy studies on Zr free, Sm(CoFe$_{0.1}$B$_{0.04}$)$_{7.5}$ as-spun ribbons at 38 m/s have shown a nonuniform microstructure, which may relate the lack of Zr to the reduced coercivity and remanence values of this sample.

## IV. CONCLUSION

Melt spinning and short-time annealing develops high coercivity in Sm–Co 2 : 17 alloys with small additions of B and Zr. The coercivity is associated to a nanocomposite microstructure, which is completely different from the one of precipitation hardened Sm$_2$Co$_{17}$ magnets with the well-known cellular/lamellar features. Exchange coupling between nanograins is responsible for loop squareness and high remanence. The nanograins of fcc–Co do not deteriorate magnetic properties when a uniform microstructure is developed. Besides boron, the presence of zirconium, up to 8% at, is also required for the development of a more uniform microstructure.

A value of 10.7 MGOe for the maximum energy product at room temperature and the good temperature variation of coercivity are very promising for the optimization of these materials and for their potential use at high temperatures.

## ACKNOWLEDGMENT

S. S. Makridis, G. Litsardakis, and D. Niarchos would like to thank Dr. N. Dempsey and Dr. D. Givord of Louis Néel Laboratory, CNRS, Grenoble, France, for kindly providing access to the 7 T extraction magnetometer.